\documentclass[amsfonts,twocolumn,prl,aps,showpacs]{revtex4}
\usepackage{graphicx}

\def\bra#1{\right\langle#1\left|}
\def\ket#1{\left|#1\right\rangle}

\begin{document}

\title{Experimental demonstration of a technique to generate arbitrary quantum superposition states}

\author{A. Ben-Kish}
\altaffiliation[Present Address: ]{Dept. of Physics, Technion,
Haifa ISRAEL}
\author{B. DeMarco}
\author{V. Meyer}
\author{M. Rowe}
\altaffiliation[Present Address: ]{NIST Optoelectronics
Division}
\author{J. Britton}
\author{W.M. Itano}
\author{B.M. Jelenkovi\'{c}}
\author{C. Langer}
\author{D. Leibfried}
\author{T. Rosenband}
\author{D.J. Wineland}
\address{NIST Boulder, Time and Frequency Division, Ion Storage Group}
\date{\today}

\begin{abstract}
Using a single, harmonically trapped $^9$Be$^+$ ion, we
experimentally demonstrate a technique for generation of arbitrary
states of a two-level particle confined by a harmonic potential.
Rather than engineering a single Hamiltonian that evolves the
system to a desired final sate, we implement a technique that
applies a sequence of simple operations to synthesize the state.
\end{abstract}
\pacs{03.67.Lx, 32.80.Qk}

\maketitle

The goal of deterministically synthesizing or ``engineering"
arbitrary states of a quantum system is at the heart of such
diverse fields as quantum computation \cite{1} and reaction
control in chemistry \cite{2}.  For harmonic oscillator states,
particular non-linear interactions can be used to generate special
states such as squeezed states. However, it is intractable to
realize a single interaction required to create an arbitrary
state. Law and Eberly \cite{3} have devised a technique for
arbitrary harmonic oscillator state generation that couples the
oscillator to a two-level atomic or ``spin" system and applies a
sequence of operations that use simple interactions.  We
demonstrate this technique on the harmonic motion of a single
trapped $^9$Be$^+$ ion and include the generation of arbitrary
spin-oscillator states \cite{4}. Such quantum state control is
relevant to the scheme for constructing a quantum computer using
trapped atomic ions \cite{7,8}, where we must control the
quantized micro-mechanical system composed of the collective ion
normal modes that are used as a data bus to transfer information
between the ion qubits. These techniques could also be used to
create input states for quantum computing schemes that use
continuous variables \cite{25} including the code-words that are
required for fault tolerant computation \cite{26}.

Arbitrary quantum state synthesis is difficult unless certain
conditions are met.  As an example, consider a simple quantum
system with four energy eigenstates labelled
$\left|0\right\rangle$, $\left|1\right\rangle$,
$\left|2\right\rangle$, and $\left|3\right\rangle$. If the system
is initially prepared in $\left|0\right\rangle$, and if couplings
that create superpositions \mbox{$\alpha_i\left
|0\rangle+\beta_i\right|i\rangle$} ($i$ = 1,2,3) can selectively
be turned on, then we can create arbitrary superpositions of the
form
\mbox{$c_0\left|0\right\rangle+c_1\left|1\right\rangle+c_2\left|2\right\rangle
+c_3\left|3\right\rangle$}.  Here, the $c_i$ are complex and
subject to the usual normalization condition
\mbox{$\sum_i\left|c_i\right|^2=1$}. This method could be realized
in an atomic system if the four states were non-degenerate levels
with different energy separations and coherent transitions
\mbox{$\left|0\right\rangle\leftrightarrow\left|i\right\rangle$}
could be driven by applied radiation. These requirements are not
often met in practice.  For example, it may be impossible to
realize all of the desired couplings
\mbox{$\left|0\right\rangle\leftrightarrow\left|i\right\rangle$}.
Also, if the eigenstates are equally spaced like the first four
energy levels of a harmonic oscillator, then driving the
\mbox{$\left|0\right\rangle\leftrightarrow\left|1\right\rangle$}
transition also induces successive transitions
\mbox{$\left|1\right\rangle\leftrightarrow\left|2\right\rangle$},
\mbox{$\left|2\right\rangle\leftrightarrow\left|3\right\rangle$},
etc. leading to fixed relations between the $c_i$.

It has long been recognized that certain interactions can cause
harmonic oscillators to evolve to particular desired states
\cite{9}. For example, if the oscillator is excited from its
ground state with the nonlinear force $F_0z\cos(2\omega t)$ then a
``vacuum-squeezed" state is created. Such states can be used to
increase measurement precision in specific applications such as
interferometry \cite{10}. However, it is usually intractable to
find the desired force or interaction that will create a state
with arbitrary coefficients.  To circumvent this problem, schemes
have been proposed \cite{11,12} that sequentially couple atomic
superposition states to the field of a cavity mode and
statistically prepare arbitrary field states through projective
measurements.  An alternative method has been proposed to
deterministically map a previously prepared superposition of
atomic Zeeman states onto the field of a cavity \cite{13}.  A more
general deterministic scheme to prepare arbitrary field states has
been suggested by Law and Eberly \cite{3,14}. The idea relies on
coupling the harmonic oscillator to an auxiliary two-level quantum
system through a sequence of simple interactions.

Consider an auxiliary system consisting of two internal states of
an atom which we label $\left|\downarrow\right\rangle$ and
$\left|\uparrow\right\rangle$ in analogy with the two-level system
resulting from a spin-$1/2$ magnetic moment in a magnetic field.
In practice, the harmonic oscillator could correspond to a single
mode of the radiation field \cite{3} or the mechanical oscillation
of a trapped atom \cite{4,15}.  The combined energy levels for
this system are depicted in Fig. 1.  As an example, we summarize
the procedure to create the state
\mbox{$\ket{\downarrow}\sum_{i=0...3}c_{\downarrow i}\ket{i}$}
starting from the ground state
$\left|\downarrow\right\rangle\left|0\right\rangle$. It is
simplest to first think about solving the inverse problem
\cite{3}: creating the state
$\left|\downarrow\right\rangle\left|0\right\rangle$ from the
initial state \mbox{$\ket{\downarrow}\sum_{i=0...3}c_{\downarrow
i}\ket{i}$}.  The procedure begins by applying a resonant pulse of
radiation that carries out a ``$\pi$-pulse"
\mbox{$\left|\downarrow\right\rangle\left|3\right\rangle\rightarrow\left|\uparrow\right\rangle\left|2\right\rangle$}
leaving no amplitude in the
$\left|\downarrow\right\rangle\left|3\right\rangle$ state
(``clearing it out") and placing amplitude $c_{\downarrow3}$ in
the $\left|\uparrow\right\rangle\left|2\right\rangle$ state.
Applying this radiation also causes transitions between other
states $\left|\downarrow\right\rangle\left|n\right\rangle$ and
$\left|\uparrow\right\rangle\left|n-1\right\rangle$. Because in
general the coherent transition rates (Rabi rates) are not the
same for different values of $n$ \cite{18}, in this first step the
amplitudes of the other states change according to
\mbox{$c_{\downarrow n}\ket{\downarrow}\ket{n}\rightarrow
c'_{\downarrow
n}\ket{\downarrow}\ket{n}+d'_{\uparrow,n-1}\ket{\uparrow}\ket{n-1}$}.
The Law/Eberly method succeeds because the state
$\left|\downarrow\right\rangle\left|0\right\rangle$ remains
unaffected since the state
$\left|\uparrow\right\rangle\left|-1\right\rangle$ does not exist.

The second step is to induce the transition \mbox{$c_{\downarrow
3}\ket{\uparrow}\ket{2}+c'_{\downarrow2}\ket{\downarrow}\ket{2}\rightarrow
c''_{\downarrow2}\ket{\downarrow}\ket{2}$}, thereby clearing out
the $\ket{\uparrow}\ket{2}$ state. The duration and phase of the
second pulse are chosen according to the known values of
$c_{\downarrow3}$ and $c'_{\downarrow2}$ in order to collapse the
superposition state.  The first two steps have cleared out the
$\ket{\downarrow}\ket{3}$ and $\ket{\uparrow}\ket{2}$ states but,
in general, non-zero amplitudes remain in the
$\ket{\uparrow}\ket{0}$, $\ket{\uparrow}\ket{1}$,
$\ket{\downarrow}\ket{0}$, $\ket{\downarrow}\ket{1}$, and
$\ket{\downarrow}\ket{2}$ states. However, by repeating this
two-step clearing-out process for successively lower values of
$n$, the state amplitudes are transferred down the dual ladder of
states eventually to the ground state $\ket{\downarrow}\ket{0}$.
Finally, to achieve the original goal, we apply these same steps
in a time-reversed fashion to carry out the mapping to
\mbox{$\ket{\downarrow}\sum_{i=0...3}c_{\downarrow i}\ket{i}$}. In
the experiments described below, we demonstrate the Law/Eberly
technique by implementing the mapping
\mbox{$\ket{\downarrow}\ket{0}\rightarrow\ket{\downarrow}\left(\ket{0}+\ket{3}\right)$}
and other intermediate mappings of the form
\mbox{$\ket{\downarrow}\ket{0}\rightarrow\sum_i\left(c_{\downarrow
i}\ket{\downarrow}+c_{\uparrow i}\ket{\uparrow}\right)\ket{i}$}
\cite{4}.

\begin{figure}[h]
\includegraphics[scale=0.45]{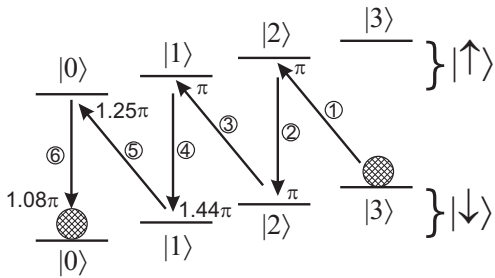}
\caption{Schematic energy level diagram for the combined harmonic
oscillator/spin-1/2 system (only harmonic oscillator levels with
$n\leq3$ are shown).  The arrows show the laser pulse sequence
used to generate the state $\ket{\downarrow}\ket{0}$ from the
state
$\ket{\Psi_{03}}=\ket{\downarrow}\left(\ket{0}+\ket{3}\right)/\sqrt{2}$.
The laser pulses are applied in a stepwise fashion (labelled from
1 to 6) where the pulse areas $\Omega_{n,n'}t$ (marked next to the
head of each arrow) are calculated according to the Rabi rate for
the numbered transitions.  In this notation, a ``$\pi$-pulse"
would completely transfer all of the population from an initial
state $\ket{\downarrow}\ket{n}$ to a final state
$\ket{\uparrow}\ket{n'}$, for example.  To generate
$\ket{\Psi_{03}}$ from $\ket{\downarrow}\ket{0}$, the pulse
sequence shown here is applied in a time-reversed
manner.\label{fig1}}
\end{figure}

The harmonic oscillator and auxiliary levels in our experiment
correspond to the motional and internal states of a single
$^9$Be$^+$ atomic ion trapped in a linear Paul trap \cite{16}.  We
use the harmonic oscillator motional states along the trap axis
($z$ direction) which are equally spaced in energy by
$h\times(2.9$~MHz), where $h$ is Planck's constant.  In this
direction, the ion is confined by a static electric harmonic
potential.  The $\ket{\uparrow}$ and $\ket{\downarrow}$
(auxiliary) spin states are the \mbox{$F=1,m_F=-1$} and
\mbox{$F=2,m_F=-2$} hyperfine levels of the ion's $^2S_{1/2}$
electronic ground state, which are separated in energy by
approximately $h\times(1250$~MHz). Applied laser radiation is used
for state preparation and manipulation.  A pair of laser beams
detuned by approximately +80~GHz from the $^2S_{1/2}$ to
$^2P_{1/2}$ electronic transition ($\lambda\sim313$~nm) drives
coherent Raman transitions and couples the $\ket{\uparrow}$ and
$\ket{\downarrow}$ states and motional levels \cite{8,17}. Motion
sensitive coupling is produced using non-collinear beams with a
wavevector difference along $z$. The Raman laser beam frequency
difference is tuned to drive \mbox{$\ket{\uparrow}\ket{n-\Delta
n}\leftrightarrow\ket{\downarrow}\ket{n}$} or
\mbox{$\ket{\uparrow}\ket{n}\leftrightarrow\ket{\downarrow}\ket{n}$}
transitions, and the coherent transition rate, or Rabi frequency,
depends on both $n$ and $\Delta n$ \cite{18,19}. The experimental
observable is the atomic spin state which we detect through
state-dependent resonance fluorescence measurements at the end of
every experiment \cite{8,20}.

To demonstrate the Law/Eberly scheme, we configure the apparatus
to generate the state
\mbox{$\ket{\Psi_{03}}=\ket{\downarrow}\left(\ket{0}+\ket{3}\right)/\sqrt{2}$}
from $\ket{\downarrow}\ket{0}$ using only transitions
\mbox{$\ket{\uparrow}\ket{n-\Delta
n}\leftrightarrow\ket{\downarrow}\ket{n}$} where $\Delta n$
alternates between 0 and 1. The ion is initialized in the
$\ket{\downarrow}\ket{0}$ state with greater than 99.9\%
probability using stimulated Raman cooling and optical pumping
\cite{21}. The six steps required to carry out the reverse process
(produce $\ket{\downarrow}\ket{0}$ from $\ket{\Psi_{03}}$) are
calculated according to the step-wise algorithm and are shown in
Fig. 1.

The state created after applying the Law/Eberly scheme is analyzed
through measurements of Rabi oscillations on the
\mbox{$\ket{\downarrow}\ket{n}\leftrightarrow\ket{\uparrow}\ket{n+\Delta
n}$}, $\Delta n$=0,$\pm1$ transitions \cite{17,23}. The
probability \mbox{$P_\downarrow=\sum_i\left|c_{\downarrow
i}\right|^2$} to detect the ion in the $\ket{\downarrow}$ state is
recorded after applying a laser pulse on one of these transitions
for duration $t$. The observed oscillations (see Fig. 3) of
$P_\downarrow$ as a function of the laser pulse duration are fit
to a sum of cosine functions with Rabi frequencies
$\Omega_{n,n+\Delta n}(=\Omega_{n+\Delta n,n})$ constrained by the
measured ratio of Rabi frequencies for the different motional
levels \cite{fitnote}.  The amplitude and phase (left as free
parameters in the fit) of each frequency component are used to
determine the probabilities $\left|c_{\uparrow i}\right|^2$ and
$\left|c_{\downarrow i}\right|^2$ for $i=0,1,2,3$ \cite{fitnote2}.
We find that the observed ion population corresponds to the target
state with 0.89 probability (Table 1), and that the populations in
$\ket{\downarrow}\ket{0}$ and $\ket{\downarrow}\ket{3}$ are equal
within the 0.03 measurement uncertainty.

\begin{figure}[h]
\includegraphics[scale=1]{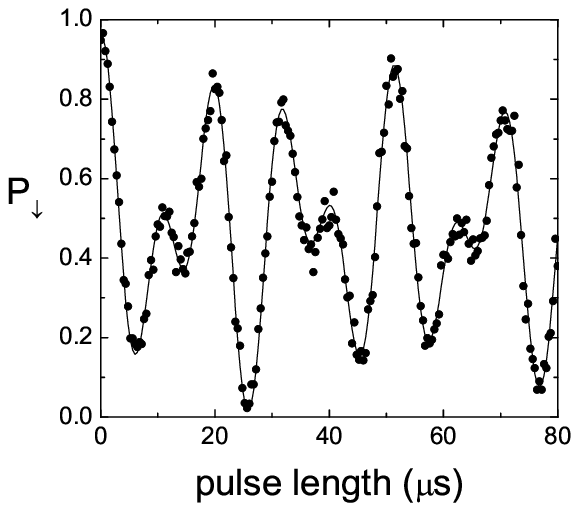}
\caption{Measured Rabi oscillations for the target state
$\ket{\Psi_{03}}$. The probability to measure the atom in the
$\ket{\downarrow}$ state is determined after applying a laser
pulse coupling states
\mbox{$\ket{\downarrow}\ket{n}\leftrightarrow\ket{\uparrow}\ket{n+1}$}
for a variable length of time. Each data point (solid circles)
represents the average of 600 experiments. The solid line is a fit
to the data that is used to determine the populations in the first
four motional states and the two spin states.  Typically the fit
determines that the $1/e$ time constant for the exponentially
decaying envelope included in the fit corresponds to 9
oscillations for the
$\ket{\downarrow}\ket{0}\leftrightarrow\ket{\uparrow}\ket{1}$
transition. The observed beating arises primarily from the
oscillations of population in the $\ket{\downarrow}\ket{0}$ and
$\ket{\downarrow}\ket{3}$ states, which have Rabi frequencies such
that $\Omega_{34}/\Omega_{01}=0.60$. The uncertainty in the spin
state discrimination is smaller than the scatter in the data,
which is mainly due to laser intensity and magnetic field
fluctuations.\label{fig2}}

\end{figure}
\begin{table}[h]
\begin{tabular}{c|c|c|c|c||}
 & n=0 & 1 & 2 & 3 \\ \hline\hline
 \parbox[c][12pt][c]{12pt}{\large $\downarrow$} & 0.43 & 0 & 0.01 & 0.46 \\ \hline
 \parbox[c][12pt][c]{12pt}{\large $\uparrow$} & 0.03 & 0.04 & 0.02 & 0.01 \\ \hline \hline
\end{tabular}
\caption{Measured state populations for the experiment with the
target state $\ket{\Psi_{03}}$.  Data similar to that in Fig. 2
are used to determine the probability $P=|c_{m_s,n}|^2$ to find
the ion in the motional level $n$ and the spin state
$m_s=\downarrow,\uparrow$ for the intended target state
$\ket{\Psi_{03}}$. This table shows the average of populations
determined from using Rabi oscillation measurements employing
couplings with $\Delta n=0,\pm1$. The uncertainty in the measured
probabilities is 0.03 and is dominated by scatter in the Rabi
oscillation data and the finite observation time. \label{fig4}}
\end{table}

The probabilities $\left|c_{\uparrow i}\right|^2$ and
$\left|c_{\downarrow i}\right|^2$ are also measured after each
step in the procedure to generate $\ket{\Psi_{03}}$ and are
compared to the theoretical predictions in Fig. 3.  The Hilbert
space trajectory from the initial to final state is somewhat
complicated, with probability appearing, at least temporarily, in
the $\ket{\downarrow}\ket{n=0,1,2,3}$ and
$\ket{\uparrow}\ket{n=0,1,2}$ states.

\begin{figure}[h]
\includegraphics[scale=0.8]{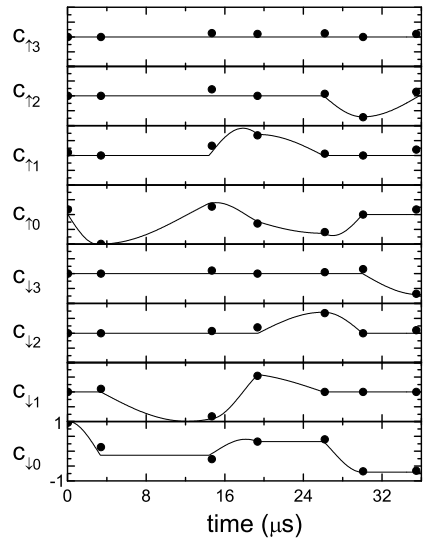}
\caption{Initial to final state Hilbert space trajectory.  The
real amplitudes $c_{\downarrow n}$ and $c_{\uparrow n}$ are shown
for the sequence used to generate $\ket{\Psi_{03}}$. The solid
lines are the theoretical prediction, while the data (solid
circles) are derived from the probabilities $|c_{\downarrow i}|^2$
and $|c_{\uparrow i}|^2$ measured using the Rabi oscillation
technique after each step. The amplitude was determined by taking
the square root of the measured probability and assigning a sign
consistent with the final state and the length of the laser
pulses.\label{fig3}}
\end{figure}

The Rabi oscillation diagnostic determines the populations
$\left|c_{\uparrow i}\right|^2$ and $\left|c_{\downarrow
i}\right|^2$, but gives no information about the phase relation
between the states.  For example, measuring the populations in
this way cannot distinguish between the pure (coherent
superposition) state described by the density matrix
\mbox{$\rho=\left(\ket{\downarrow}\ket{0}+\ket{\downarrow}\ket{3}\right)\left(\bra{0}\bra{\downarrow}+\bra{3}\bra{\downarrow}\right)/2$}
and the mixed (incoherent) state described by
\mbox{$\rho=\left(\ket{\downarrow}\ket{0}\bra{0}\bra{\downarrow}+\ket{\downarrow}\ket{3}\bra{3}\bra{\downarrow}\right)/2$}.
To verify that our implementation of the Law/Eberly scheme
establishes coherence we have performed a test experiment starting
from $\ket{\downarrow}\ket{0}$ using the target state
\mbox{$\ket{\Psi_T}=0.64\ket{\downarrow}\ket{0}+0.77\ket{\uparrow}\ket{2}$},
which would give $\left|c_{\downarrow0}\right|^2=0.41$ and
$\left|c_{\uparrow2}\right|^2=0.59$. The first five pulses of the
$\ket{\Psi_{03}}$ sequence were used to generate $\ket{\Psi_T}$.
The measured probabilities for the experimentally generated state
were $\left|c_{\downarrow0}\right|^2=0.39$,
$\left|c_{\uparrow2}\right|^2=0.55$,
$\left|c_{\uparrow0}\right|^2=0.03$, and 0.03 distributed among
the remaining states.

A coherent analysis pulse was applied on
\mbox{$\ket{\uparrow}\ket{n+2}\leftrightarrow\ket{\downarrow}\ket{n}$}
transitions after the $\ket{\Psi_T}$ state generation pulses but
before spin state detection. The laser pulse area was adjusted to
be a ``$\pi/2$"-pulse for the
\mbox{$\ket{\uparrow}\ket{2}\leftrightarrow\ket{\downarrow}\ket{0}$}
transition. For the pure $\ket{\Psi_T}$ state, the population
would almost fully oscillate between the states
$\ket{\downarrow}\ket{0}$ and $\ket{\uparrow}\ket{2}$ as the phase
of the analysis pulse was varied relative to the state generation
pulses.  No sensitivity to this phase would be observed if the
state we generated was an incoherent mixture of populations
(dashed line in Fig. 4).  The measured probability to find the
atom in the state $\ket{\downarrow}$ as the laser phase is swept
is shown in Fig. 4. The amplitude of these oscillations can be
related to the fidelity
\begin{eqnarray}
F&=&\langle\Psi_T|\rho\ket{\Psi_T}\nonumber\\
&=&0.41\rho_{\downarrow0\downarrow0}+0.59\rho_{\uparrow2\uparrow2}+0.495\left(\rho_{\downarrow0\uparrow2}+\rho_{\uparrow2\downarrow0}\right)\nonumber
\end{eqnarray}
where $\rho$ is the experimentally measured density matrix. We
determine that $F=0.93\pm0.03$ using the measured populations and
oscillation contrast to determine the relevant elements of the
density matrix $\rho$ as in ref \cite{Sackett00}.

\begin{figure}[h]
\includegraphics[scale=0.8]{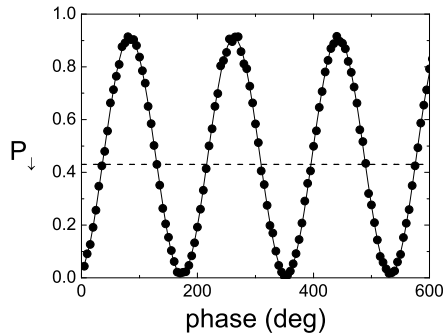}
\caption{Coherence fringes.  After generation pulses for the
target state
\mbox{$0.64\ket{\downarrow}\ket{0}+0.77\ket{\uparrow}\ket{2}$} are
implemented, a ``$\pi/2$" analysis laser pulse is applied with a
controlled phase relative to the state generation pulses.  Shown
here are the resulting oscillations in the probability
$P_\downarrow$ to find the atom in the $\ket{\downarrow}$ state as
the laser pulse phase is swept. The amplitude of the oscillation
determined from a fit to a cosine function (solid line) is used to
establish the fidelity of the experimentally generated state.  The
oscillation centered around $P_\downarrow=0.46$ is consistent with
the measured probability in the $\ket{\uparrow}\ket{0}$ state
(which is unaffected by the analysis laser coupling) and
experimental error in the analysis laser pulse duration.  The
dashed line indicates the result if the prepared state is an
incoherent mixture.\label{fig5}}
\end{figure}

In summary, we have demonstrated experimentally the scheme of Law
and Eberly \cite{3} for generation of arbitrary
harmonic-oscillator states and its extension \cite{4} to arbitrary
harmonic-oscillator/spin states. The method can be generalized to
higher dimensions \cite{4}, to the generation of arbitrary density
matrices of harmonic oscillators \cite{14}, to the creation of
arbitrary motional observables \cite{15} such as the phase
\cite{24}, and to the generation of arbitrary Zeeman state
superpositions \cite{29}. The precision with which we can
implement this technique has a direct relation to the efficiency
of quantum-information processing using trapped ions \cite{7}.
With sufficient improvements in the fidelity of such operations,
one can contemplate using additional motional modes of motion as
information carriers in this scheme. Of course, the same
techniques can be applied in cavity-QED, the system in which it
was originally conceived \cite{3}. More generally, such techniques
increase the variety of tools available for quantum-information
processing and may eventually find application in areas not
currently anticipated.

We thank D. Lucas and J. Ye for helpful comments on the
manuscript.  This work was supported by the NSA and the ARDA under
Contract No. MOD-7171.00. Contribution of the U.S. Government: not
subject to U.S. copyright.


\begin{references}
\bibitem{fitnote} We include the possibility for population only
in the $n=0,1,2,3$ levels in the fits.  We estimate that the
probability that ambient heating and off-resonant coupling to
other sideband transitions populates levels higher than $n=3$
after the state generation pulses is less than 0.1\%.
\bibitem{fitnote2} The fit assumes phases of only 0 or $\pi$
between coherent superposition amplitudes that remain between
coupled states.  If other phases exist, the Rabi oscillation
contrast is reduced and we underestimate the populations.  We
determine that, to a good approximation, other phases can be ruled
out because the sum of populations (left as a free parameter) sum
to unity within the fit uncertainty.
\end{references}
\end{document}